\documentclass[prd,a4paper,notitlepage,twocolumn,superscriptaddress,10pt]{revtex4}
\usepackage{amssymb,amsfonts,amsmath,graphicx,color,mathptm,times,amssymb,hyperref}
\usepackage[centering,hmargin=2cm,vmargin=2.5cm]{geometry}
\usepackage[utf8]{inputenc}
\usepackage{soul}
\usepackage{hyperref}
\include{hl}
\usepackage[T1]{fontenc}
\usepackage{currvita}

\begin{document}
\date{\today}
\title{On the weight of entanglement}

\author{David Edward Bruschi\footnote{Current affiliation:York Centre for Quantum Technologies, Department of Physics, University of York, Heslington, YO10 5DD York, UK}}
\address{Racah Institute of Physics and Quantum Information Science Centre, the Hebrew University of Jerusalem, 91904 Givat Ram, Jerusalem, Israel}
\email{david.edward.bruschi@gmail.com}

\begin{abstract}
We investigate a scenario where quantum correlations affect the
gravitational field. We show that quantum correlations between
particles occupying different positions have an effect on the
gravitational field. We find that the small perturbations induced by
the entanglement depend on the amount of entanglement and vanish for
vanishing quantum correlations. Our results suggest that there is a
form of entanglement that has a weight, since it affects the
gravitational field. This conclusion may lead towards a new
understanding of the role of quantum correlations within the overlap of
relativistic and quantum theories.
\end{abstract}

\maketitle

\section{Introduction}

\textit{Does entanglement have a weight?} A positive answer to this
question would have far reaching consequences, since entanglement is
the core resource of some of the most exciting applications of the
field of quantum information. For example, entanglement can be used for
teleportation \cite{Bennett:Brassard:93}, quantum key distribution
\cite{Hoi-Kwong:Marcos:14} and quantum computing \cite
{Ladd:Jelezko:10} to name a few. More importantly, a positive answer
would also help us deepen our understanding of the overlap of
relativistic and quantum theories.

Quantum entanglement is a type of correlation that, to date, is not
known to interact with gravity. The role of quantum correlations in
gravitational scenarios has been so far ignored, most likely due to the
fact that overwhelming experimental evidence shows that entanglement
can be well established between different systems in the presence of a
gravitational field apparently without noticeable consequences \cite
{Ma:Herbst:12}. However, experiments are reaching regimes where small
modifications introduced by the mutual effects of entanglement and
gravity might be measured \cite{Bruschi:Ralph:2014,Vallone:Bacco:14}.
Therefore, in the last decade attention has been given to understand
the effects of gravity on entanglement \cite{Alsing:Fuentes:12}. Most
approaches indicate that effects of gravity on entanglement should
exist, although we lack the theory of quantum gravity that can
naturally predict this. Unfortunately, the effects predicted by this
body of work do not arise because of a direct coupling between gravity
and quantum correlations. In particular, it cannot be shown that
entanglement will affect gravity, the necessary step to conclude that
gravity and entanglement interact with each other.

In this work, we establish that quantum correlations affect the
gravitational field and that small perturbations in the metric are
induced by the presence of quantum coherence. We employ Einstein's
equations and semiclassical theory to show that, for low energy (few
particle) states, a small control parameter naturally arises and is
uniquely determined by the energy scales of the problem.
We then find that small changes in the metric depend on the amount of
entanglement present in the state, as measured by the logarithmic
negativity (and, additionally, by the concurrence), and vanish for
vanishing quantum correlations. These effects are ``radiated away'' for
times larger than the decoherence time, which we show is proportional
to the ``size'' of the particle. Furthermore, the relative phase of the
coherences has a direct influence on the magnitude of the effects. Our
results are complementary to previous work which investigated the
stability of coherent superpositions of different energy states in the
presence of gravity \cite{Penrose:96}. They are also related, for
example, to previous work that investigated spontaneous collapse of the
wave function due to gravity
\cite{Bassi:Lochan:13}, to stochastic gravity \cite{Hu:Verdaguer:08}
and the role of coherent superpositions \cite{Ford:Roman:2008}.
However, contrary to most of this body of work, we are not interested
here in the effects of gravity on quantum states (i.e., collapse of the
wave function) but rather on the back-reaction of quantum coherence on
gravity. Finally, we argue that the regimes considered in this work are
well within the limits of validity of semiclassical gravity \cite
{Phillips:Hu:2000,Anderson:Molina:2003}.

We believe that our results have important implications for both
quantum and relativistic theories, in particular they aid theoretical
and experimental research to look for phenomena which might challenge
our current understanding of nature.

\section{Background}

\subsection{Quantum field theory in curved spacetime}

In this work particles are excitations of quantum fields that propagate
on a classical spacetime. We consider for simplicity a massive scalar
quantum field $\phi(x^{\mu})$ with mass $m$ in \mbox{$(3+1)$}-dimen\-sional
spacetime \cite{BandD} with metric $g_{\mu\nu}$ (see \cite
{conventions}). The equation of motion of the field is $(\square
+m^2)\phi=0$, where the d'Alambertian is $\square\equiv(\sqrt
{-g})^{-1} \partial_{\mu} [\sqrt{-g} g^{\mu\nu} \partial_{\nu}]$
(a standard reference is \cite{BandD}).

The field can be decomposed in any orthonormal basis of solutions
$u_{\mathbf{k}} (x^{\mu})$ to the Klein--Gordon equation as $\phi
=\int
d^3k[a_{\mathbf{k}} u_{\mathbf{k}} + a^{\dagger}_{\mathbf{k}}
u^*_{\mathbf{k}}]$, with annihilation and creation operators
$a_{\mathbf{k}}$, $a^{\dagger}_{\mathbf{k}}$ that satisfy the canonical
commutation relations
$[a_{\mathbf{k}},a^{\dagger}_{\mathbf{k}'}]=\delta^3({\mathbf
{k}}-{\mathbf{k}}')$
and all other vanish. The annihilation operators $a_{\mathbf{k}}$
define the vacuum state $\vert0\rangle$ through $a_{\mathbf{k}}
\vert0\rangle=0\  \forall\, \mathbf{k}$. In general, it is
convenient to
choose the set of modes $\{u_{\mathbf{k}}\}$ if it satisfies (at least
asymptotically) an eigenvalue equation of the form $i\,\partial_\tau
u_{\mathbf{k}}=\omega\,u_{\mathbf{k}}$, where $\partial_\tau$ is some
(possibly global) time-like Killing vector and
$\omega:=\omega_{\mathbf{k}}=\sqrt{{\mathbf{k}}\cdot{\mathbf{k}}+m^2}$
is a real eigenvalue \cite{BandD}.

We assume that the spacetime is essentially flat Minkowski with metric
$g_{\mu\nu}=\eta_{\mu\nu}=\mathrm{diag}(-1,1,1,1)$ and perturb the
flat spacetime metric in the following way
%
\begin{align} \label{metric:expansion}
g_{\mu\nu}=\eta_{\mu\nu}+\xi\,h_{\mu\nu},
\end{align}
where we have introduced the small control parameter $\xi$ i.e., $\xi
\ll1$. In this work we will consider only effects that are
proportional to $\xi$ i.e., we ignore $\mathcal{O}(\xi^2)$
contributions. Here $h_{\mu\nu}$ depends on the spacetime coordinates
$x^{\sigma}$ and evolves dynamically via Einstein equations. The
expansion  {\eqref{metric:expansion}} is known as linearised (or linear)
gravity, which has been successfully employed, for example, to predict
the existence of gravitational waves \cite{Sathyaprakash:Schutz:09}.
The role of the parameter $\xi$ is pivotal and we will show in the
following that it is uniquely determined by the relevant physical
energy scales.

\subsection{Semiclassical gravity}

We wish to take into account the back reaction of the quantum field on
the metric, in other words, we wish to take into account the fact that
a single excitation of the field is responsible for the perturbation
$h_{\mu\nu}$. This can be done within the framework of semiclassical
gravity, which has been successfully applied \cite
{Hartle:77,Albrecht:Steinhardt:82,Anderson:83} but has its own domain
of validity \cite{Phillips:Hu:2000,Anderson:Molina:2003}. Since our
work involves only considering mean energy, which is a quantity that
can be experimentally measured, the scenario considered in this work
falls within this domain of validity and we will comment on this later
\cite{Ford:Roman:2008}.

In this framework, back reaction is implemented through Einstein's
semiclassical equations
%
\begin{align} \label{semiclassical:einstein:equation}
G_{\mu\nu}=-8\,\pi\,G_N\,\langle T_{\mu\nu}\rangle_{\mathrm{Ren}},
\end{align}
where $G_{\mu\nu}$ is Einstein's tensor, $G_N$ is Newton's constant,
$T_{\mu\nu}$ is the stress energy tensor of the quantum field and
``Ren'' stands for some choice of renormalisation of the stress energy
tensor \cite{BandD}. The average $\langle\cdot\rangle$ is intended
over some chosen initial state $\rho$ of the field. Einstein's tensor
contains second derivatives of the metric, which account for its
dynamics. However, one needs to be careful with correctly identifying
the source of the gravitational field, a process called
renormalisation. That care needs to be taken into account in curved
spacetime is a well known issue \cite{BandD}. Many methods have been
proposed and employed with different degrees of success \cite
{Christensen:76,Wald:77}. However, in this work we will analyse the
back reaction of single particle excitations on flat Minkowski
spacetime. We believe it is natural to assume that in this case it is
sufficient to subtract the (infinite) zero point energy of the
Minkowski vacuum, a procedure known as normal ordering \cite{BandD}.
We therefore have $\langle T_{\mu\nu}\rangle_{\mathrm{Ren}}\equiv
\langle:T_{\mu\nu}:\rangle$, where the symbol $:\cdot:$ stands for
normal ordering \cite{BandD}.

The metric is coupled to the field via the semiclassical Einstein
equation  {\eqref{semiclassical:einstein:equation}}. In order to exploit
this relation we need to compute stress energy tensor $T_{\mu\nu}$
which is readily found in literature \cite{BandD} as $T_{\mu\nu
}=\partial_{\mu}\phi\partial_{\nu}\phi-\frac{1}{2}\,g_{\mu\nu
}[\partial^{\rho}\phi\partial_{\rho}\phi-m^2\phi^2]$. The field
$\phi$ satisfies the equation of motion $(\square+m^2)\phi=0$ with
the full metric $g_{\mu\nu}$ in~ {\eqref{metric:expansion}}.

Since we choose to look at effects at lowest order in $\xi$ we can
expand the field as $\phi=\phi^{(0)}+\xi\,\phi^{(1)}$, where $\phi
^{(0)}$ satisfies $(\partial^{\rho}\partial_{\rho}+m^2)\phi
^{(0)}=0$ defined with the flat metric $\eta_{\mu\nu}$. We therefore
find that $\phi^{(0)}=\int d^3k[a_{\mathbf{k}} u_{\mathbf{k}} +
a^{\dagger}_{\mathbf{k}} u^*_{\mathbf{k}}]$, where the plane wave
modes $u_{\mathbf{k}}$ take the form $u_{\mathbf{k}}=(2\,\pi\,
)^{-3/2}\,\omega^{-1/2}\,\exp[i\,k_{\mu}\,x^{\mu}]$ and $k_{\mu}\,
x^{\mu}=-\omega\,t+\mathbf{k}\cdot\mathbf{x}$. The correction
$\phi^{(1)}$ to the field satisfies a more complicated differential
equation but turns out to be irrelevant for our purposes. Furthermore,
we notice that the state does not depend directly on $\xi$. It will
become evident that the parameter $\xi$ appears in the right hand side
of  {\eqref{semiclassical:einstein:equation}} only through the
\textit{average} of the stress energy tensor over the initial state.

By considering a small coupling to gravity it is easy to show that the
first order contributions to the semiclassical Einstein equation  {\eqref{semiclassical:einstein:equation}} satisfy the following equation
%
\begin{align} \label{first:order:semiclassical:einstein:equation}
\tilde{G}^{(1)}_{\mu\nu}=-8\,\pi\,\langle:\tilde{T}^{(0)}_{\mu
\nu}:\rangle,
\end{align}
where the dimensionless tensors $\tilde{G}^{(1)}_{\mu\nu}$ and
$\tilde{T}^{(0)}_{\mu\nu}$ are obtained from the dimensional
Einstein tensor $G_{\mu\nu}$ and stress energy tensor $T_{\mu\nu}$
respectively.

\subsection{Relevant initial states and physical control parameters}
\label{sec:twomodecosmo}

We now wish to determine the parameter $\xi$ in terms of the relevant
(energy) scales of the problem. We choose to work in the Heisenberg
picture and will analyse the following two-parameter family of initial states
%
\begin{align} \label{initial:states}
\rho(\alpha,\beta)=&\,\alpha\,|01\rangle\langle01|+(1-\alpha)\,
|10\rangle\langle10|+\beta\,|10\rangle\langle01|\nonumber\\
&+\beta\,|01\rangle\langle10|,
\end{align}
where $0\leq\alpha\leq1$, the parameter $-1/2\leq\beta\leq1/2$ is
real without loss of generality and $(\alpha-1/2)^2+\beta^2\leq1/4$
in order for $\rho(\alpha,\beta)$ to represent a physical state.
Notice that for $\alpha=1/2$ and $\beta=0$ one has a maximally mixed
state while for $\alpha=\beta=1/2$ one has a maximally entangled
state. Furthermore, we underline that the sign of the parameter $\beta
$ might play an important role in the final effects and we will comment
on this later.

We note here that the state  {\eqref{initial:states}} is the most general
state that obeys the ``superselection rule'' greatly discussed in
literature \cite{Penrose:96}, i.e., that it is not possible to
superpose states with different masses. In fact, the terms in  {\eqref{initial:states}} are the only one particle state terms that have the
same (average) energy. Terms such as $|0\rangle\langle0|$ or
$|11\rangle\langle11|$ clearly have a different (average) energy.

Here we define the normalised single particle states $|01\rangle$ and
$|10\rangle$ as excitations over the Minkowski vacuum of the
\textit{same} particle in \textit{different} positions in the following way
%
\begin{align} \label{initial:particle:states}
|01\rangle:=&\int d^3k\, F_{\mathbf{k}_0}(\mathbf{k})\,e^{-i\,
\mathbf{L}\cdot\mathbf{k}}\,a^{\dag}_{\mathbf{k}}|0\rangle
\nonumber\\
|10\rangle:=&\int d^3k\, F_{\mathbf{k}_0}(\mathbf{k})\,e^{i\,\mathbf
{L}\cdot\mathbf{k}}\,a^{\dag}_{\mathbf{k}}|0\rangle,
\end{align}
where we have introduced the peaked functions $F_{\mathbf
{k}_0}(\mathbf{k})$, the constant $\mathbf{k}_0$ defines the location
of the peak in momentum space (aligned along the $z$-direction without
loss of generality, i.e. $\mathbf{k}_0=(0,0,k_0)$), the vector $\pm
\mathbf{L}$ defines the location of the peak in position space which
are located at a distance of $2L:=2\,\sqrt{\mathbf{L}\cdot\mathbf
{L}}$, (again, along the $z$-direction without loss of generality,
$\mathbf{L}=(0,0,L)$). The creation operators $a^{\dag}_{\mathbf
{k}}$ are the flat spacetime Minkowski operators associated with the
zero order field $\phi^{(0)}$. Furthermore, normalisation implies that
$\int d^3k\,|F_{\mathbf{k}_0}(\mathbf{k})|^2=1$.

We need to make sure that the particle states  {\eqref{initial:states}}
are orthogonal (at least to good approximation) in order for the
concept of entanglement between the two excitations to have a proper
meaning. We can choose between two different profile functions. One
choice is a Gaussian profile function $F_{\mathbf{k}_0}(\mathbf
{k})=(\sqrt[4]{8\,\pi^2\,\sigma^2}\,\sigma)^{-1}\,\exp[-\frac
{(\mathbf{k}-\mathbf{k}_0)^2}{4\,\sigma^2}]$, where $\sigma$ is the
width of the profile and is assumed to be large, which makes the
excitation very localised in position space. We can compute the overlap
of the particle states and find $|\langle10|01\rangle|=|\int d^3k\,
|F_{\mathbf{k}_0}(\mathbf{k})|^2\,\exp[-2\,i\,\mathbf{L}\cdot\nobreak
\mathbf{k}]|\propto\allowbreak\,\exp[-2\,\sigma^2\,L^2]$ which is negligible
for large separations compared to the spread of the wave packet i.e.,
$\sigma\,L\gg1$. This choice might lead to problems when one wishes
to look at states with higher numbers of particles. In that case, the
overlap of the new states can become larger, which might lead to
question the meaning of the following work. We therefore turn to a
second choice, the box profile function i.e., $F^{\prime}_{\mathbf
{k}_0}(\mathbf{k})=(\sqrt{8\,\sigma}\sigma)^{-1}\operatorname{Rect}(\frac{k_x-k_{0,x}}{2\,\sigma})\operatorname{Rect}(\frac
{k_y-k_{0,y}}{2\,\sigma})\operatorname{Rect}(\frac{k_z-k_{0,z}}{2\,\sigma
})$ where $\mathrm{Rect}(x)$ is the rectangle function. In this case, we
can choose $\mathbf{L}=(0,0,n\,\pi/\sigma)$ with $n\in\mathbb{Z}$
which guarantees
orthogonality between the particle states i.e., $\langle10|01\rangle
\equiv\,0$.

We then notice that the parameter $\sigma$ acts as a natural scale for
energies (or equivalently lengths in natural units). In order to
understand the interplay of the energy scales of the problem we
introduce the dimensionless wave numbers $\tilde{\mathbf{k}}:=\mathbf
{k}/\sigma$ and the dimensionless coordinates $\tilde{x}^{\mu
}:=\sigma\,x^{\mu}$. We then notice that Einstein's tensor $G_{\mu
\nu}$ to first order can be written as a combination of second
derivatives of the metric. We can therefore introduce $\tilde{G}_{\mu
\nu}:=G_{\mu\nu}/\sigma^2$, where the dimensionless tensor $\tilde
{G}_{\mu\nu}=\xi\,\tilde{G}^{(1)}_{\mu\nu}$ appears in  {\eqref{first:order:semiclassical:einstein:equation}}. Without loss of
generality and to obtain analytical results, we focus on two
interesting regimes for the field excitations: that of extremely
massive static particles ($m/\sigma\gg1$ and $k_0=0$) and that of
 massless particles with high momentum ($m=0$ and $k_0/\sigma\gg1$). It
follows that the average of the stress energy tensor of a particle
excitation will be, to good approximation, proportional to $m\,\sigma
^3$ or $k_0\,\sigma^3$ respectively (this can be found from a
straightforward computation of stress energy tensor components i.e.,
$\langle:T^{(0)}_{00}:\rangle$). We therefore have $\langle
:T^{(0)}_{\mu\nu}:\rangle=E_0\,\sigma^3\, \langle:\tilde
{T}^{(0)}_{\mu\nu}:\rangle$, where $E_0$ is proportional to $m$ or
$k_0$ depending on the regime. Putting all together in  {\eqref{semiclassical:einstein:equation}} we have
%
\begin{align}\label{intermediate:semiclassical:einstein:equation}
\xi\,\tilde{G}^{(1)}_{\mu\nu}=-8\,\pi\,G_N\,E_0\,\sigma\,\langle
:\tilde{T}^{(0)}_{\mu\nu}:\rangle+\mathcal{O}(G_N\,E_0\,\sigma\,
\xi).
\end{align}
We conclude that  {\eqref{intermediate:semiclassical:einstein:equation}}
identifies $\xi=G_N\,E_0\,\sigma\ll1$ and confirms that  {\eqref{first:order:semiclassical:einstein:equation}} holds to lowest order.
Higher order terms on the right hand side contain first order
correction to the stress energy tensor and do not contribute to the
effects of interest here. However, effects to this order would include
the direct coupling of quantum correlations with gravity.

\section{Interplay of gravity and entanglement}

\subsection{First order contribution to the curvature}

We now proceed to outline our main results. The semiclassical Einstein
equation  {\eqref{first:order:semiclassical:einstein:equation}} for the
initial state $\rho(\alpha,\beta)$ is
%
\begin{align} \label{semiclassical:einstein:equation:explicit}
\tilde{G}^{(1)}_{\mu\nu}=&\alpha\,\langle01|:\tilde{T}^{(0)}_{\mu
\nu}:|01\rangle+(1-\alpha)\,\langle10|:\tilde{T}^{(0)}_{\mu\nu
}:|10\rangle\nonumber\\
&+2\,\beta\,\Re\,(\langle01|:\tilde{T}^{(0)}_{\mu\nu}:|10\rangle).
\end{align}
We conclude from  {\eqref{semiclassical:einstein:equation:explicit}} that
the Einstein tensor $G^{(1)}_{\mu\nu}$ has a contribution that comes
purely from quantum coherence. The term $\beta\,\Re\,(\langle
01|:\tilde{T}^{(0)}_{\mu\nu}:|10\rangle)$ is responsible for such
difference and its contribution to the metric is therefore proportional
to $\beta$. We quantify the entanglement present in the state $\rho
(\alpha,\beta)$ by employing the logarithmic negativity $E_{\mathcal
{N}}$ which is bound by $0\leq E_{\mathcal{N}}\leq1$ (see \cite
{Adenaert:Plenio:03}). This is a well known measure of entanglement and
is defined as $E_{\mathcal{N}}=\log_2(2\,\mathcal{N}+1)$, where the
negativity $\mathcal{N}$ is defined as $\mathcal{N}=\sum_n(|\lambda
_n|-\lambda_n)/2$ and $\lambda_n$ are the eigenvalues of the partial
transpose of the state $\rho(\alpha,\beta)$. We find that $|\beta
|=(2^{E_{\mathcal{N}}}-1)/2$ which proves that the last term in  {\eqref{semiclassical:einstein:equation:explicit}} contributes only when there
are some quantum correlations i.e., $E_{\mathcal{N}}\neq0$. The
greatest contribution from this term occurs when $E_{\mathcal{N}}=1$
i.e., $\alpha=\beta=1/2$ and the state $\rho(1/2,1/2)$ is maximally
entangled.

Finally, given that  {\eqref{initial:states}} is the most general state
we can consider, and that the system effectively behaves as a system of
two qubits, we are in the position of computing the concurrence $C$ for
this system and, if desired, the entanglement of formation $E_{oF}$
\cite{Wootters:1998}. The entanglement of formation, which can be
computed in our case as a simple function of the concurrence, captures
all of the correlations and enjoys an important information-theoretical
and practical interpretation: it quantifies the minimum number of
copies of maximally entangled states of qubits necessary to prepare the
state with only Local Operations and Classical Communications (LOCC)
\cite{Wootters:1998}. For states like ours, the concurrence has been
already computed and has the simple expression $C=2\,|\beta|$. This
corroborates our claim that quantum correlations are responsible for
the effects described in this work.

We could now proceed to compute all (ten independent) terms in  {\eqref{semiclassical:einstein:equation:explicit}}. This can be done
explicitly however, since the main aim of this work is to show that an
effect exists in the first place, we find it more convenient to compute
the Ricci scalar $R:=-G^{\mu}{}_{\mu}$ which gives a more compact
result and measures the strength of the curvature locally at each
point. To achieve this goal, we note that it is sufficient to compute
$\mathcal{D}^{01}{}_{\mu}{}^{\mu}:=\langle01|:\tilde
{T}^{(0)}{}_{\mu}{}^{\mu}:|01\rangle$ (or equivalently any other of
the terms) since all other terms can be obtained by $\mathcal
{D}^{01}{}_{\mu}{}^{\mu}$ with simple modifications.
We find
%
\begin{align} \label{expectation:value:zero:one}
\mathcal{D}^{01}{}_{\mu}{}^{\mu}=&\,{-}\frac{1}{\sigma^2}\int d^3k\,
d^3k'\,e^{i\,\mathbf{L}\cdot(\mathbf{k}'-\mathbf{k})}\,F_{\mathbf
{k}_0}(\mathbf{k})F_{\mathbf{k}_0}(\mathbf{k}')\nonumber\\
&\,\times\left[k_{\mu}'\,k^{\mu}+2\,m^2\right]u^{*}_{\mathbf
{k}}u_{\mathbf{k}'}.
\end{align}
It is straightforward to show that the other diagonal term\break $\mathcal
{D}^{10}{}_{\mu}{}^{\mu}:=\langle10|:\tilde{T}^{(0)}{}_{\mu
}{}^{\mu}:|10\rangle$ can be obtained from
 {\eqref{expectation:value:zero:one}} by replacing $\mathbf{L}\rightarrow
-\mathbf{L}$ in the integrand and the off diagonal term $\mathcal
{D}^{0110}{}_{\mu}{}^{\mu}:=\langle01|:\tilde{T}^{(0)}{}_{\mu
}{}^{\mu}:|10\rangle$ can be obtained from
 {\eqref{expectation:value:zero:one}} by replacing
$\tilde{\mathbf{k}}-\tilde
{\mathbf{k}}'\rightarrow\tilde{\mathbf{k}}+\tilde{\mathbf{k}}'$
in the exponent inside the integrand.

We continue by discussing the contribution of all these terms to the
time evolution of the curvature (i.e., Ricci scalar). We start by
noticing that all terms contain a factor of the form $\exp[\pm\,i\,
(\omega-\nobreak \omega')t]$. When $\sigma^2\hbar\,t/m\gg1$ for extremely
massive fields, or $\sigma\,t\gg1$ for massless fields, all terms on
the right hand side of  {\eqref{semiclassical:einstein:equation:explicit}} vanish due to
Riemann--Lebesgue lemma. We understand this is a consequence of the
spreading of the wave packets $F_{\mathbf{k}_0}(\mathbf{k})$ with
time \cite{note}.

Focusing on the initial time $t=0$, one can show that the term
$\mathcal{D}^{10}{}_{\mu}{}^{\mu}$ has the expansion $\mathcal
{D}^{10}(x,y,z)+\mathcal{O}((\frac{\sigma}{m})^2)$ for massive
static particles and $\mathcal{D}^{10}(x,y,z)+\mathcal{O}(\frac
{\sigma}{k_0})$ for massless particles with high momentum.
It is possible to compute the function $\mathcal{D}^{10}(x,y,z)$ for
the box wave-packets and we find $\mathcal{D}^{10}(x,y,z)\sim
\mathrm{sinc}^2(\sigma\,x)\operatorname{sinc}^2(\sigma\,y)\operatorname{sinc}^2(\sigma
\,(z-L))$. In this case, the term\break $\mathcal{D}^{01}(x,y,z)$ can be
found by the previous one by replacing $z-L$ with $z+L$ and the term
$\mathcal{D}^{0110}(x,y,z)$ by replacing $\mathrm{sinc}^2(\sigma\,
(z-L))$ with $\sin^2(\sigma\,z)/(L^2-z^2)$. Note that here we have
used the fact that $\sin(L\,\sigma)=\sin(\pi\,n)=0$.

The off diagonal terms $\mathcal{D}^{0110}{}_{\mu}{}^{\mu}$ do
contribute to Einstein's tensor in the fashion described above and to
the same order in $\sigma/m$ or $\sigma/k_0$ as the diagonal terms.
If one is interested in obtaining the metric itself, one can integrate
equation  {\eqref{semiclassical:einstein:equation:explicit}} with similar
contributions as determined above and obtain the form of the
perturbation $h_{\mu\nu}$ for all states, which we have shown will
depend on $\alpha$ and $\beta$. This could be done numerically
however, we are not interested in doing so here, as the aim of this
work is to prove that an effect exists in the first place.

\subsection{Physical regimes}

We have shown that correlations affect gravity and that, for small
perturbations of flat spacetime, the coupling is governed by the
dimensionless parameter $\xi$. Furthermore, this parameter is fully
determined by the relevant physical scales of the scenario i.e., energy
scales. Let us now restore dimensions in order to understand which is
the magnitude of the effects governed by $\xi$ and the time $\tau$ it
takes for the gravitational field to completely ``wash out'' all the effects.
We start by looking at the control parameter $\xi$. We have
%
\begin{align} \label{control:parameter:with:dimensional:constants}
\xi=\frac{G_N\,E_0\,\sigma}{c^4},
\end{align}
where we have noted that $E_0=m\,c^2$ for massive static particles and
$E_0=\hbar\,k_0\,c$ for massless particles with high momentum. For a single
massive particle whose rest mass $m\sim10^{-21}\mbox{\ kg}$ is much
larger its
``size'', of the order of $1/\sigma\sim10^{-22}\mbox{\ m}$ (see
\cite{Dehmelt:88}), we see that $\xi\sim10^{-26}$. For a single
massless particle with high momentum
(frequency) $\omega_0\sim
10^{14}\mbox{\ Hz}$ compared to its spread $\sigma\,c\sim10^9\mbox
{\ Hz}$ we find
$\xi\sim10^{-63}$, which is extremely small. However, for much heavier
particles, for ultra-energetic massless particles or for states with a
high number of excitations (i.e., N00N states, which have already been
employed to greatly enhance estimation of parameters due to their
``high'' quantum nature \cite{Kok:Lee:02}), one could hope to increase
the above result by several orders of magnitude. This could in
principle make the effect measurable.

We notice that, for a very massive and static particle, the parameter
$\xi$ can be re-written as $\xi=r_S/r$, where
$r_S:=\frac{2\,G_N\,m}{c^2}$ is the Schwarzschild radius of a particle
of mass $m$ and ``size'' $r=2/\sigma$. The predictions of this work
become unreliable when the Schwarzschild radius of the particle becomes
comparable and exceeds the size of the particle.

Let us turn to the time $\tau$ it takes for these effects to become
negligible. We have seen that the components of Einstein's tensor
vanish after times that depend on the particle being massive ($\tau_m$)
or massless ($\tau_{k_0}$). In particular
%
\begin{align} \label{decoherence:time}
\tau_m:=\frac{m}{\sigma^2\,\hbar}, \quad\tau_{k_0}:=\frac
{1}{\sigma\,c}.
\end{align}
Given the numbers considered above we have $\tau_m\sim10^{-32}\mbox
{\ s}$ and
$\tau_{k_0}\sim10^{-31}\mbox{\ s}$ respectively. A possible way to
increase the
lifetime of the contributions would be to consider particles that have
very well defined momentum i.e., lower $\sigma$.

Surprisingly, it appears that the sign of $\beta$ affects the results
and can make the final effect (slightly) bigger or smaller. This can be
generalised to complex $\beta$. Furthermore, notice that although the
timescale is independent of the parameter $\beta$, the vanishing
effect occurs equally to all for first order contributions. There is no
such behaviour for zero order contributions to the off diagonal terms. Finally, we notice that the magnitude of the effects, or changes, depends on the amount of entanglement (on the absolute value of $\beta$), while the direction of the contribution (an increase or decrease) depends on the phase (equivalently, the sign of $\beta$).

We now comment on the consistency of the methods and the results. It
has been argued that criteria for the validity of semiclassical gravity
should depend on the state considered and on the scales probed \cite
{Phillips:Hu:2000,Anderson:Molina:2003}. In particular it has been
showed that, for Minkowski space and lower than Planck scales \cite
{Anderson:Molina:2003} and smeared fields (as the ones considered here)
which do not probe scales much smaller than the smearing size \cite
{Phillips:Hu:2000}, the semiclassical treatment is valid and should
give correct predictions.
As a consistency check on the results, we note that if $E_0=0$ or
$G_N=0$ the effects described in this work vanish. This is to be
expected since in this case there would be no excitations to produce
the perturbation of the metric or no dynamical gravity.

\subsection{Considerations on the scope and validity of the results}

A few final comments are in place. First, we have analysed states that
do not have coherent superpositions or mixtures of single particle
states with \textit{different} mass (energy). This property is crucial
to our results. On the one hand, our results are not affected by
arguments that suggest that gravity should collapse states that are
coherent superpositions of states with different energy (in line with
\cite{Penrose:96}). On the other, it guarantees that the states  {\eqref{initial:states}} are the most general one particle states that we can
consider. This in turns guarantees that entanglement is directly responsible for the effects described in this work. Second, we note that not all entanglement affects gravity.
For example, we could look at states of particles entangled in the spin
degree of freedom. In the absence of magnetic fields, spin up and spin
down are both eigenstates of the same hamiltonian operator (i.e., the
energy levels are degenerate in the spin degree of freedom). In this
case, entanglement between spins would not interact with gravity.
Third, it may be tempting to draw an analogy between the semiclassical
equations used here and, for example, semiclassical electromagnetism.
One might seek for a direct analogy between equation  {\eqref{semiclassical:einstein:equation}} and, for example,
%
\begin{align}
\partial_{\mu}F^{\mu\nu}=\frac{\mu_0\,q}{c}\langle:J^{\nu
}:\rangle,
\end{align}
where $F^{\mu\nu}:=\partial^{\mu}A^{\nu}-\partial^{\nu}A^{\mu}$
is the classical Faraday tensor, $A^{\mu}$ is the classical
four-vector potential, $J^{\nu}:=-i[\phi\,\partial^{\nu}\phi^{\dag
}-\partial^{\nu}\phi\,\phi^{\dag}]$ is the current of the now
charged scalar field $\phi$, the constant $\mu_0$ is the magnetic
permeability of the vacuum and $q$ is the charge of the field
excitations. In the same fashion as done in this work, one seeks to
expand four potential and current as $A_{\mu}=A_{\mu}^{(0)}+\xi\,
A_{\mu}^{(1)}$ and $J_{\mu}=J_{\mu}^{(0)}+\xi\,J_{\mu}^{(1)}$
respectively, where $\xi\ll1$ is a parameter to be determined. Note
that, in order to compare with the gravitational case, we consider a
perturbation of the vector potential around the zero order $A_{\nu
}^{(0)}$ which satisfies the \textit{homogenous} Maxwell equation $\partial
_{\mu}\partial^{\mu}A_{\nu}^{(0)}-\partial_{\mu}\partial^{\nu
}A_{\mu}^{(0)}=0$. This allows us to compare this scenario with the
gravitational case, where the zero order component of the metric (i.e.,
the Minkowski metric $\eta_{\mu\nu}$) satisfies $G_{\mu\nu}=0$.
One then looks for the dimensionless version of $\partial_{\mu}F^{\mu
\nu}=\frac{\mu_0\,q}{c}\langle:J^{\nu}:\rangle$ and wishes to
obtain the analogous of equation  {\eqref{first:order:semiclassical:einstein:equation}}. However, since both
vector potential and current are dimensional, after simple algebra one
finds $\xi\,[\partial^{\mu}\partial_{\mu}\tilde{A}_{\nu
}^{(1)}-\partial^{\mu}\partial_{\nu}\tilde{A}_{\mu
}^{(1)}]=\langle:\tilde{J}_{\nu}^{(0)}:\rangle$. Here quantities
with a tilde are dimensionless and the derivates are with respect to a
normalised coordinate.
The expansion parameter $\xi$ is arbitrary and is not fixed by the
physics of the problem. Furthermore $\tilde{A}_{\nu}^{(1)}$ and
$\langle:\tilde{J}_{\nu}^{(0)}:\rangle$ are independent of $\xi$.
Therefore, the relation $\xi\,[\partial^{\mu}\partial_{\mu}\tilde
{A}_{\nu}^{(1)}-\partial^{\mu}\partial_{\nu}\tilde{A}_{\mu
}^{(1)}]=\langle:\tilde{J}_{\nu}^{(0)}:\rangle$ cannot be satisfied
and this perturbative expansion is inconsistent.
We conclude that, although the main equations of these two
semiclassical theories are formally similar, the physics they describe
are essentially different and cannot be compared. We understand that
this difference is a consequence of the universality of gravity, which
couples to all energy, while the electromagnetic field couples only to charge.

Finally, our results suggest that entanglement is responsible for the
effects described in this work. The initial state  {\eqref{initial:states}} is the most general one-particle state that can be
conceived given the constraint that superpositions of different masses
(or energies) are not allowed \cite{Penrose:96}. Entanglement in this
state is always present when off diagonal terms are, which corroborates
our claims.

\section{Discussion}
To summarize, we have shown that entanglement can affect the
gravitational field. This suggests that entanglement ``has a weight''.
The perturbations in the gravitational field depend on the amount of
entanglement and vanish for vanishing quantum correlations.
The effects studied in this work decay with a time scale proportional
to the characteristic ``size'' of the particle but that does not depend
on the amount of entanglement. Furthermore, relative phase of the
coherence term seems to directly affect the strength of the effect.
A prospective theory of quantum gravity must be able to account for
this phenomenon and explain its origin.

Experiments designed to measure these effects will have to carefully
balance the different parameters, in particular the distance at which
the entanglement is established and the energy of the particle. We
believe that our results can help in better understanding the overlap
of relativity and gravity theories and, ultimately, in the quest of a
theory of quantum gravity.

\section*{Acknowledgments}
We thank Marcus Huber, Leila Khouri, Johannes Niediek, Dennis R\"atzel,
Bei-Lok Hu, Paul R. Anderson, Larry Ford, \v{C}aslav Brukner and Marco
Piani for useful suggestions and discussions. We extend special thanks
to the late Jacob Bekenstein for very insightful comments and suggestions, to
Jorma Louko for extremely valuable correspondence on details
of the results of this work, to Ivette Fuentes for helping strengthen
the results of this work and to Gerard Milburn for illuminating
discussions on possible ways to detect the effects.
D. E. B. was supported by the I-CORE Program of the Planning and Budgeting Committee and the Israel Science Foundation (grant No. 1937/12), as well as by the Israel Science Foundation personal grant No. 24/12.

We would like to dedicate this work to the memory of Prof. Jacob
Bekenstein, inspiring and visionary pioneer of black hole physics and
quantum gravity.

\bibliographystyle{unsrt}
\bibliography{WeightPaper}

\begin{thebibliography}{10}
\bibitem{conventions}
In this paper we will use the natural convention $c=\hbar=k_B=1$ and we use
  Einstein's summation convention. The metric has signature $(-,+,+,+)$.
\bibitem{note}
We note that in $2+1$ dimensions there are contributions to Einstein's tensor
  that do not decay in time i.e., those for which $|\mathbf{x}|\sim t$.
\end{thebibliography}

\end{document}